\documentclass[letterpaper,9pt]{article}
\usepackage{fullpage}
\usepackage{amssymb}
\usepackage{amsbsy}
\usepackage{amsfonts}
\usepackage{amsmath}
\usepackage{dsfont}
\usepackage{epsfig}
\usepackage{subfigure}
\usepackage{graphicx}
\usepackage{epstopdf}

\def\bea{\begin{eqnarray}}
\def\eea{\end{eqnarray}}
\def\beqn{\begin{eqnarray}}
\def\eeqn{\end{eqnarray}}
\def\beq{\begin{equation}}
\def\eeq{\end{equation}}

\def\Dslash{\not{\hbox{\kern-4pt $D$}}}
\def\pslash{\not{\hbox{\kern-4pt $p$}}}

\begin{document}

\begin{titlepage}

\setcounter{page}{0}
\begin{flushright}

\end{flushright}
\vskip 3cm

\begin{center}

{\huge\bf Making the Sneutrino a Higgs with a $U(1)_R$ Lepton Number }

\vskip 1cm

{\large {\bf Claudia Frugiuele} and {\bf Thomas Gr\'egoire}}

\vskip 0.5cm
{\it Ottawa-Carleton Institue for Physics , Department of Physics,  Carleton University, \\
               1125 Colonel By Drive, Ottawa, Canada, K1S 5B6} \\

\abstract{We present a supersymmetric extension of the Standard Model (SM) that posseses a continuous $ U(1)_R$ symmetry, which is identified with one of three lepton numbers, and where a sneutrino vev gives mass to 
the down type quark and leptons.
This idea allows for a smaller particle content than the minimal $R$-symmetric supersymmetry extension of the standard model (MRSSM).  We explore bounds on this model coming from electroweak precision measurements, neutrino masses and gravitino decay. Bounds from electroweak precision measurements lead to a two-sided bound on $\tan \beta$ while gravitino decay forces a low reheating temperature. Finally, the generation of neutrino masses from $R$-symmetry violation put an upper bound on the SUSY breaking scale. Despite all of this, we find that the allowed parameter space is still large and would lead to a distinctive phenomenology at the LHC.

}

\end{center}
\end{titlepage}

\section{Introduction}
Supersymmetry (SUSY) at the weak scale remains one of the favorite paradigm for physics at the terascale. In the minimal supersymmetric version of the Standard Model (MSSM) and many extensions thereof, the weak scale is protected against large quadratically divergent radiative corrections, there exist a natural dark matter candidate and gauge couplings unify at a high scale. Unfortunately the fact that LEP, Tevatron and now the first data from LHC did not find any superpartners or the Higgs makes the realization of this scenario difficult without a fair amount of fine-tuning. This motivates the exploration of a larger portion of the weak scale supersymmetry landscape. For example,  one can consider models where the gaugino soft masses are Dirac instead of Majorana \cite{Hall:1990hq,Randall:1992cq,Fox:2002bu}. This requires the introduction of new superfields and a different couplings of the susy breaking sector to the Standard Model gauge sector. These soft Dirac gaugino masses do not contribute to the running of scalar soft masses, and are therefore dubbed 'supersoft' \cite{Fox:2002bu}. This could help create a small hierarchy between gaugino and scalar masses which might be an interesting starting point in order to improve on the fine tuning issues of the MSSM. It also allow the possibility  of writing models that are invariant under a full $U(1)_R$ instead of the usual $R$-parity. The flavour constraints on such models are relaxed and supersymmetry breaking can be transmitted to the visible sector through gravitational interactions \cite{Kribs:2010md}
. Because $U(1)_R$ symmetry forbid a $\mu$ term, the Higgs sector of these models need to be different than the MSSM. One option \cite{Kribs:2007ac} is to enlarge the field content and include two new doublets $R_u$ and $R_d$. The other option  \cite{Davies:2011mp} is to give masses to the down-type quarks and to the leptons through a  SUSY-breaking term . 

In this work we examine the possibility of instead giving masses to the down-type quarks and to the leptons through the vev of a sneutrino (the idea of giving the sneutrino a vev has a long history, see \cite{sneutrinovev} for examples). In the MSSM, the lepton doublet has the same quantum number as the down type Higgs, it can therefore serve this purpose. However, in the MSSM such a vev  is very strongly constrained, mainly due to the fact that it breaks lepton number and  induce neutrino masses that are too large. In models with a $U(1)_R$ symmetry however, the $U(1)_R$ can be identified with a lepton number (see \cite{fayet} for an early implementation of this idea), and the sneutrino can acquire a relatively large vev \cite{Gherghetta:2003wm}. The goal of this paper is to explore the main features of such a scenario. In section \ref{Themodel} we present the particle content of the model and the Lagrangian. Because one of the lepton number is a $U(1)_R$ symmetry, the gauginos  carry a lepton number and mix the corresponding lepton and neutrino. Constraints on such mixing from electroweak precision measurement are presented in section \ref{constraints}.  In the same section we present constraints that arise from gravitino decay, and also from the generation of neutrino masses through unavoidable $R$-symmetry breaking. In section \ref{Rsusybreaking} we discuss possibilities for mediating SUSY breaking in such a model and the related $\mu/B_\mu$ problem. Finally in section \ref{pheno} we discuss the main features of the collider phenomenology.

\section{ \label{Themodel}The model}
\subsection{Particle content and Lagrangian}
The particle content of our model consists of the usual particle content of the MSSM to which we add an adjoint chiral superfield $\Phi_i$   
 for each SM gauge group  $G_i= SU(3)_C,SU(2)_L, U(1)_Y$. This is  necessary to give  Dirac mass to the gauginos and is the minimal particle content needed to accommodate a $U(1)_R$ symmetry in a supersymmetric extension of the Standard Model.  In fact, this particle content is more minimal than the minimal R-symmetric supersymmetric extension of the Standard Model (MRSSM) presented in \cite{Kribs:2007ac}, as the latter includes two additional weak doublets in order to give mass to the gauginos as the standard $\mu$ term is forbidden by $R$-symmetry. We therefore refer to our model as the MMRSSM.
 Table  \ref{table:particlecontent}  shows the MMRSSM superfields and their quantum numbers; 
 the $ R$ charge assignments  is chosen such that we can use the $R$-symmetry as the lepton number of type $a$, where $a=e,\mu$ or  $\tau$. Indeed all the Standard Model particles, except the charged lepton $ a^-$ and the neutrino $\nu_a,$ carry $R$-charge zero.
The situation with the SUSY partners is  reversed: the charged slepton and the sneutrino of flavour $a$ do not carry any lepton number while all other have lepton number. This means in particular that a sneutrino vev does not break the lepton number,
 and this is  crucial for  making the sneutrino the down type Higgs. The squarks are leptoquarks because they carry both the baryon number and the lepton number $a.$ As we will show in a following section, this feature  characterizes and distinguishes the phenomenology of the model.
Moreover,  the higgsinos, the  wino, the bino together with their adjoint partners carry $ R $-charge $ \pm 1;$ this means they can mix with the ordinary leptons of flavor $a.$ In the MMRSSM the lightest chargino and the lightest neutralino coincide with the charged lepton $ a^{ \pm}$ and the neutrino $ \nu_a.$
\par
 The up-type Higgs $ H_u$  has $R$ charge 0, and it acquires a vev. Instead, $R_d $ is an inert doublet\footnote{ This is a common feature of our model with the SOHDM \cite{Davies:2011mp}.}, which  is introduced to cancel the $H_u$ anomalies, and
to give mass to the higgsinos. It is the sneutrino of flavor $a$ that acquires a vev and gives mass to the down-type fermions.
\begin{table}
\centering 
{
\begin{tabular}{l|l|l} 
\multicolumn{1}{c|}{\textbf{SuperField}} & 
\multicolumn{1}{c|}{\textbf{$ ( SU(3)_c,SU(2)_L)_{U(1)_Y} $}} & 
\multicolumn{1}{c}{\textbf{$U(1)_R$}} \\ 
\hline 
$Q_i$ & \  \ \ \ \ \ \ $ (3,2)_{\frac{1}{6}}$ \ \ & 1 \\ 
$U^c_i$ & \  \ \ \ \ \ \ $ (\bar 3,1)_{-\frac{2}{3}}$ \ \ & 1 \\  
$D^c_i$ & \  \ \ \ \ \ \ $ (\bar 3,1)_{\frac{1}{3}}$ \ \ & 1 \\ 
$E^c_a$ & \  \ \ \ \ \ \ $ (1,1)_{1}$ \ \ & 2 \\ 
$L_a$ & \  \ \ \ \ \ \ $ (1,2)_{-\frac{1}{2}}$ \ \ & 0 \\ 
$E^c_{b,c}$ & \  \ \ \ \ \ \ $ (1,1)_{1}$ \ \ & 1 \\ 
$L_{b,c}$ & \  \ \ \ \ \ \ $ (1,2)_{-\frac{1}{2}}$ \ \ & 1 \\ 
$H_u$ & \  \ \ \ \ \ \ $ (1,2)_{\frac{1}{2}}$ \ \ & 0 \\
$R_d$ & \  \ \ \ \ \ \ $ (1,2)_{-\frac{1}{2}}$ \ \ & 2 \\ 
$ \Phi_{\tilde W} $ &  \  \ \ \ \ \ \ $ (1,1)_0$ & 0 \\ 
$ \Phi_{\tilde B} $ &  \  \ \ \ \ \ \ $ (1,3)_0$ & 0 \\ 
$ \Phi_{\tilde g}$ &  \  \ \ \ \ \ \ $ (8,1)_0$ & 0\\

\end{tabular} 
}\qquad\qquad 

\caption{R-charge assignment for the chiral supermultiplets in our model. The subscript $a$ denote the flavour of the lepton superfield that plays the role of the down-type Higgs. The subscrit $b,c$ represent the remaining two flavours of leptons. } 
\label{table:particlecontent}
\end{table} 
\newline
With this particle content the MMRSSM superpotential is then:
\begin{equation} 
W= \textbf{  y}_u  U^c Q H_u-   \textbf{y}_d  D^c Q L_{a}- y_b E^c_b L_{b}  L_{a}-y_{c} E^c_c L_{c}  L_{a}
+ \mu H_u R_d.
\label{eq:superpotential}
\end{equation}
where $ \textbf{y}_u$ and $\textbf{ y}_d $ are $3 \times 3 $ matrices in family space, while $ a,b,c= e,\mu$
and $ \tau.$ 
\par
As usual, the up-type fermions acquire mass through $H_u,$  while the down type Yukawa couplings involve the leptonic superfield $L_a,$ which then plays the role of the down-type Higgs. 
However, it is important to note that  the superpotential  in equation \eqref {eq:superpotential} 
does not contain the Yukawa coupling for the lepton of flavor $ a $ as the term $ L_{a} L_{a} E_a^c $ is null, while the term $ R_d  L_{a}  E_a^c $  is forbidden by  the $R$-symmetry.
Therefore, this coupling needs to be generated in the SUSY breaking sector as  we will discuss in a following section.
The down-type Yukawa couplings of equation \eqref {eq:superpotential}  violate the conventional $R_p$ parity   as well as the standard lepton number.
Indeed, here  these  couplings 
 correspond to the  trilinear $ R_p$ violating coupling  $ \lambda_{ijj} L_i L_j E^c_j,$ and  $ \lambda'_{ijj} L_i Q_j D^c_j$ often discussed in the literature \cite{Barbier:2004ez}.  These  couplings have very stringent bounds in conventional $R$-parity breaking models that come from the Majorana neutrino masses they induce. In our model however, there is a conserved lepton number which forbids such masses. In fact in the limit of massless neutrinos, we impose three separate lepton numbers, one for each flavour: $U(1)_{R_a}$ which is the $R$-symmetry as well as $U(1)_b$ and $U(1)_c$ which are not $R$ symmetries. As a consequence, the bounds on those coupling are in the MMRSSM much less stringent than in conventional $R$-parity violating models, and come mainly from  electroweak precision measurements.
 This, as we will see, has interesting phenomenological consequences.
\par
The inert doublet $R_d$ does not interact with the SM fermions as the trilinear couplings $ D^c Q R_d, $ and $ E^c L R_d $  are forbidden by the $R$-symmetry. 
As we have already commented, $R_d$ is necessary to give mass to the higgsinos.
Indeed, a bilinear  term $ H_u L_{a} $ is forbidden by the $R$-symmetry, and the higgsinos acquire mass through the $R$-symmetric  $ \mu $ term $ H_u R_d.$ 
\par
Finally, the soft supersymmetry breaking terms allowed by both gauge symmetries and by the $R$-symmetry are:
\begin{align}
\mathcal{L}_{soft}&=  \mathcal{L}^{f}_{mass}+\mathcal{L}^{s}_{mass}
-  B_{\mu} ( H_u \tilde l_{a}+cc),
\label{eq:softsusy}
\end{align}
where the gaugino masses are given by:
\begin{equation}
\label{eq:L_f}
\mathcal{L}^f_{mass}=  M_{\tilde B} \lambda_{\tilde B} \psi_{\tilde B}+M_{\tilde W}  \lambda_{\tilde W}  \psi_{\tilde W}+M_{\tilde g} \lambda_{\tilde g} \psi_{\tilde g},
\end{equation}
and the soft scalar masses by:
\begin{align}
\label{eq:L_s}
\mathcal{L}^s_{mass}&=  m^2_{ \tilde q} \tilde q^{\dagger} \tilde q+ m^2_{  \tilde l} \tilde l^{\dagger} \tilde  l+
m^2_{ \tilde u^c} \tilde  {u^c}^{\dagger} \tilde u^c+ m^2_{ \tilde d^c} \tilde {d^c}^{\dagger} \tilde d^c \\ \nonumber
&+m^2_{e} \tilde {e^c}^{\dagger} \tilde e^c
 +m^2_{H_u} H_u^{\dagger} H_u+ m^2_{R_d} R_d^{\dagger} R_d+ m^2_{\Phi_{\tilde B}} \Phi_{\tilde B}^{\dagger} \Phi_{\tilde B} + \\ \nonumber
 & +  m^2_{\Phi_{\tilde W}} \Phi_{\tilde W}^{\dagger} \Phi_{\tilde W}+ m^2_{\Phi_{\tilde g}} \Phi_{\tilde g}^{\dagger} \Phi_{\tilde g}+ M^2_{\Phi_{\tilde B}} (\Phi_{\tilde B}^{2}+cc)+
  M^2_{\Phi_{\tilde W}} (\Phi_{\tilde W}^{2}+cc)+ M^2_{\Phi_{\tilde g}} (\Phi_{\tilde g}^{2}+cc).
\end{align}
We notice that  equation \eqref{eq:softsusy} contains a $B$-term that mixes the $\tilde\nu_a$ sneutrino with $ H_u,$ but not a mixing term for $ r_d.$ This ensures that $ r_d $ will not get a vev  as long it does not acquire a negative mass while the sneutrino will. 
Moreover, we note  that  the soft SUSY lagrangian   of equation \eqref{eq:softsusy} does not contain  scalar trilinear coupling $ A_{ijk}$ nor  Majorana mass terms for the gauginos. 
\par
As we have observed in the introduction,  $R$-symmetric  models can be generated through the supersoft SUSY breaking mechanism \cite{Fox:2002bu}.
In this scenario, supersymmetry breaking is parametrized  by a non-dynamical vector superfield with a non-zero D-term. 
The Dirac masses for the gauginos in equation \eqref{eq:L_f} are then generated by the following operator:
\begin{equation}
\int{ d^2\theta  \frac{c_i} {M} W_{\alpha}' W^{\alpha}_i \Phi_i},
\end{equation}
where $ W_{\alpha}' $ is the field strength superfield of the SUSY breaking spurion: $W_{\alpha}' = \theta_\alpha D'$.  This is known as a supersoft operator as the Dirac gaugino masses it produces will not lead to log-divergent susy breaking scalar masses. In principle it allows that gauginos to be parametrically heavier than the sfermions.  The adjoint scalars can also get a mass from an operator involving $W'$:
\begin{equation}
\int{ d^2\theta  \frac{ W'_{\alpha} W^{' \alpha} }{M^2} \Phi^2_i } \; \;.
\label{phi2}
\end{equation}
This operator gives rise to the term proportional to $M^2_{\Phi_{i}}$ in equation \eqref{eq:L_s}. It gives a positive mass square to the real part of $\Phi_i$, but a negative mass square which is potentially dangerous to the imaginary part of $\Phi_i$.  In theories of R-symmetric gauge mediation which we will consider for this model one can also generate an operator of the form \cite{Carpenter:2010as,Benakli:2008pg}:
\begin{equation}
\int d^4 \theta \frac{{W'}^\alpha D_\alpha V' \Phi_i^\dagger \Phi_i}{M^2}
\end{equation}
which gives a common mass squared$m_{\Phi_i}^2$ to the real and imaginary parts of the adjoint scalar. With appropriate choice of couplings for the messengers \cite{Benakli:2009mk}
it is possible in R-symmetric gauge mediation to avoid having tachyonic adjoint scalars. 
Moreover, they are the heaviest particle of the spectrum with their masses parametrically the square root of a loop factor above the gauginos masses.
The sfermions  are the lighest superpartners with soft masses squared that come from the following finite one loop contribution \cite{Fox:2002bu,Benakli:2009mk}: 
\begin{align}
m_s^2= \sum_{b=1}^3 \frac{C^b_s \alpha_b M^2_{b}}{\pi} \log{\frac{ M^2_{\Phi_{Rb}^2}}{M^2_b}},
\end{align}
where  $ C^b_s  $ is the quadratic Casimir of the scalar $s$ under the group $b,$ which is 
equal to $ Y^2(s) $ for $ U(1)_Y,$ and $ \frac{N^2-1}{2N} $ for $SU(N).$
\par
Finally, as we have already anticipated, the SUSY breaking lagrangian  should contain the Yukawa coupling $ y_{a} \tilde h_u^{\dagger} e^c_{a} l_a $.  This term needs to come from the mechanism of SUSY breaking mediation and we will discuss it's origin in section \ref{Rsusybreaking}
 \subsection{Electroweak symmetry breaking}
In the present section we will study how electroweak symmetry breaking is realized in our model. Such an analysis was also done for a quite general model in \cite{Belanger:2009wf} . The part of the potential that is relevant for electroweak symmetry breaking contains only $h_u^0$, $\tilde \nu_a$ as well as the adjoint scalars $ \tilde \phi_{\tilde B},$ and
 $\tilde \phi_{\tilde W} $  as they can acquire a non-zero  vev. All other fields do not get a vev and are set to $0$ in what follows. The potential consists of three terms:

\begin{align}
V_{EW}=V_{D}+V_F+V_{soft} \;\;\;.
\end{align}
The first is the contribution from the $SU(2)_L$ and $U(1)_Y$ D-term and is given by:
\begin{align} 
V_D &= \frac{1}{2} ( \sqrt{2}M^2_{\tilde B} ( \tilde \phi_{\tilde B}+ \tilde \phi_{\tilde B}^{\dagger})+ \frac{ g'}{2} ( |H^0_u|^2-|\tilde \nu_a|^2))^2+ 
 \frac{1}{2} ( \sqrt{2}M^2_{\tilde W} ( \tilde \phi^0_{\tilde W}+ \tilde \phi^{0*}_{\tilde W})+ \frac{g}{2} ( |H^0_u|^2-|\tilde \nu_a|^2))^2,
\end{align}
The second contribution comes, instead, from the superpotential, and it  only contains a mass term for the up-type Higgs:
\begin{align}
V_F= \mu^2 |H^0_u|^2.
\end{align}
Finally, the third contribution contains the following soft SUSY breaking terms:
\begin{align} 
V_{soft}&= m^2_{ \tilde \phi_{\tilde B}}  \tilde \phi_{\tilde B}^{\dagger}  \tilde \phi_{\tilde B}+ m^2_{ \tilde \phi_{\tilde W}}  \tilde \phi_{\tilde W}^{\dagger}  \tilde \phi_{\tilde W}+ M^2_{ \tilde \phi_{\tilde B}} ( \tilde \phi_{\tilde B}^{2}+cc)+ 
  M^2_{ \tilde \phi_{\tilde W}} ( \tilde \phi_{\tilde W}^{2}+cc)+  \\ \nonumber
 & m^2_{H_u} |H_u^0|^2+m^2_{L_a} |\tilde \nu_a^2|- B_{\mu} ( H^0_u \tilde \nu_a+h.c.).
\end{align}
The scalar potential is then:
\begin{align} 
V_{EW}=&  (\mu^2+ m^2_{H_u}) |H^0_u|^2+  m^2_{\tilde \nu_a} |\tilde \nu_a|^2- B_{\mu} ( H^0_u \tilde \nu_a+h.c.)+
\frac{g'+g}{8} (|H_u|^0-|\tilde \nu_a|^2)^2+ \\ \nonumber
 & + \frac{1}{2} (m^2_{ \tilde \phi_{\tilde B}}+M^2_{ \tilde \phi_{\tilde B}}+4 M^2_{\tilde B})   \tilde \phi^{R 2}_{\tilde B}+
g' M_{\tilde B}^2  \tilde \phi^R_{\tilde B} (|H_u|^0-|\tilde \nu_a|^2)+ g M_{\tilde W}^2  \tilde \phi^R_{\tilde W} (|H^0_u|^2-|\tilde \nu_a|^2).
\end{align}
with $\tilde \phi_i^R$ denoting the real part of $\tilde \phi_i$. 

%

As we have already noticed, in gauge mediation models, the adjoint scalars 
are the heaviest particle of the spectrum \cite{Benakli:2008pg} and can be integrated out of the potential. This has two effects: first it lowers the Higgs quartic and second  it shift the mass of the $Z$ boson, creating a contribution to the $\rho$ parameter:
\begin{equation}
\Delta \rho = \frac{v^2}{M^4_{\Phi^R_W}}g^2 M_{\tilde W} \cos \left(2 \beta\right) \; \;,
\end{equation}
where $ M^2_{\Phi^R_W} = m^2_{\Phi_{\tilde W}}+M^2_{\Phi_{\tilde W}}+4 M^2_{\tilde W}$ is the mass of the real part of the $SU(2)$ adjoint scalar and $\tan \beta$ is the ratio of the vev of the up-type Higgs and the vev of the sneutrino: $\tan \beta = v_u/v_a$. With $M_{\Phi^R_W}$ larger than a few TeV, the above contribution to $\rho$ is within the experimental bound, and we can neglect the correction to the Higgs potential and minimize the following potential: 
\begin{align}
V_{EW}= ( \mu^2 + m^2_{H_u}) |H_u^0 |^2 + m^2_{\tilde \nu_a}  | \tilde \nu_a|^2- B_{\mu} ( H^0_u \tilde \nu_a+h.c.)
 + \frac{ g^2+g'^2}{8} ( |H^0_u|^2-|\tilde \nu_a|^2)^2.
\end{align}
This is exactly the scalar potential of the MSSM with $ H^0_d \rightarrow  \tilde \nu_a, $ except  that here we do not have the 
$ \mu $ contribution to the  sneutrino $ \tilde \nu_a $ mass, as the  $R$ invariant $ \mu $ term contains only  $ H_u.$
Therefore, in order for the potential to be bounded from below 
the quadratic part should be positive along the $ D$ flat directions:
\begin{equation}
2B_\mu <  \mu^2 + m^2_{H_u}+ m^2_{\tilde \nu_a}.
\end{equation}
Furthermore, the condition for electroweak symmetry breaking is:
\begin{equation}
B_\mu> (\mu^2 + m^2_{H_u})  m^2_{\tilde \nu_a}.
\end{equation}
\begin{align}
\sin{\beta}&= \frac{ 2 B_{\mu}}{m^2_{H_u}+\mu^2+m^2_{L_a}} ,\\
M^2_Z&= \frac{|\mu^2+ m^2_{H_u}-m^2_{L_a}|}{\sqrt{1-\sin^2{\beta}}}-m^2_{H_u}-m^2_{L_a}-\mu^2.
\end{align}
The spectrum of the Higgs sector of the model contains the usual CP odd neutral particle $ A^0, $ the two CP even $H^0,h^0,$ and the charged Higgs. Their masses are:
\begin{align}
m^2_{A^0}&=  \frac{2 b}{ \sin{2  \beta}}=m^2_{H_u}+\mu^2+m^2_{L_a}, \\
m^2_{H^{\pm}}&= m^2_{A^0}+m^2_{W} ,  \\
m^2_{h^0,H^0}&=  \frac{1}{2}Ê(m^2_{A^0}+ m^2_{Z ^0} \mp + \sqrt{( m^2_{A^0}-m^2_{Z^0})^2 + m^2_{A^0}  m^2_{Z^0}  \sin^2{2 \beta}}.
\end{align}
This is identical to the case of the MSSM and we therefore,  we inherit also the  MSSM little hierarchy problem.
In a  $ R$-symmetric model this problem could be even more severe. The $ R$-symmetry forbids the left/right stop mixing, and this reduces the contribution of the stop radiative corrections to the SM Higgs mass.
  Indeed, the full  one loop contribution of the stop sector to the Higgs mass is \cite{Martin:1997ns}:
 \begin{align}
 \delta m^2_{h^0}& =  \frac{3}{ 4 \pi^2} \sin^2{\beta} y^2_t [ m_t^2 \ln({ \frac{ m_{\tilde t_1}m_{\tilde t_2}}{ m^2_{ t}}} )+ c^2_{\tilde t} s^2_{\tilde t}  (m^2_{\tilde t_2}-m^2_{\tilde t_1})
  \ln({ \frac{ m^2_{\tilde t_1}}{ m^2_{\tilde t_2}}} ) + \\ \nonumber
  &+ c^4_{\tilde t} s^4_{\tilde t}  ((m^2_{\tilde t_2}-m^2_{\tilde t_1}) ^2- \frac{1}{2} (m^4_{\tilde t_2}-m^4_{\tilde t_1})
  \ln({ \frac{ m^2_{\tilde t_1}}{ m^2_{\tilde t_2}}} ))/ m^2_t  ],
  \label{eq:stoprad}
 \end{align}
 where $c_{\tilde t}, $ and  $s_{\tilde t} $ are the cosine and the sine of the stop mixing angle, and $ \tilde t_1, \tilde t_2$ the mass eigenstate.  
 From equation \eqref{eq:stoprad}  we see how the absence of left/right  mixing considerably reduces the radiative contribution from the stop sector forcing  the mass of the stop to increase in order to make the Higgs sufficiently  heavy. 
 \par
 On the other hand, the supersoft SUSY breaking mechanism ameliorates the fine tuning problem, because now the radiative contribution to $M^2_{{H_u}} $ is:
 \begin{align}
\Delta M^2_{{H_u}}=\frac{ 3 y_t  m_{\tilde t}^2}{4 \pi^2} \ln{ \frac{   m_{\tilde t}}{\Lambda}},
\end{align}
 where the cutoff scale $ \Lambda $ is the mass of the real adjoint scalars, and  not the messenger scale as in the typical gauge mediation scenarios.
 \par
 As in the MSSM,  one might wonder how to increase the Higgs quartic coupling, and reduce in this way the fine tuning.
 The only  $R$ symmetric dimension five operator that  gives a contribution to the Higgs quartic coupling is:
\begin{equation}
\int{ \frac{ d^2\theta}{M} (H_u H_d) ( H_u L_a)}.
\end{equation}
A possible way  to generate this operator is to introduce a singlets  which couples to the Higgs superfields in the following way:
 \begin{equation}
m_S  S \bar S +k_1 H_u H_d S  + k_2  H_u L_a \bar S.
\end{equation}
 This is a possible solution to the little hierarchy problem in our model  inspired by the NMSSM.
Alternatively, if we consider a very low SUSY breaking scale $ \frac{F}{M^2}\sim 1$ 
we might be able to increase the Higgs quartic coupling through the following operator:
\begin{align}
\int{ d^4\theta \frac{ X^{\dagger}X}{M^4} ( H_u^{\dagger} H_u)^2}.
\end{align}
We plan to explore in more detail the fine tuning problems of the model in  future work.
\subsection{Lepton mixing}
In the MMRSSM all the sparticles are $a$ leptons, except for the sneutrino and the slepton of flavour $a.$  In particular,  the new fermions (gauginos, adjoints, higgsinos), and the neutrino $ \nu_a$  as well as the charged
lepton $ a^-$ carry $R$ charge $ \pm1,$ and therefore they can all mix.
\par
In the gauge eigenstate basis with $ \Psi_+=( \tilde W^+, \psi^+_{\tilde W}, \tilde H_u^{+},a^c)$ and $\Psi_-=( \tilde W^-,\psi^-_{\tilde W}, \tilde R_d^{-}, a^{-})$ the chargino mass term is given by:
\par
\begin{equation}
\mathcal{L}_C=\Psi_-^T  M_C \Psi_+,
\end{equation}
where:
 \[
 M_C =
\begin{pmatrix} 
0 &M_{\tilde W}& -\frac{g v_u}{\sqrt{2}} & 0 \\ 
  M_{\tilde W} &  0&0   & 0\\
0  & 0& \mu &0 \\
   -\frac{g v_{a }}{ \sqrt{2}} &  0 & 0 &m_a \\

  \end{pmatrix} \; \;.
\] 
The smallest eigenvalue corresponds to the mass of the charged lepton $ a^-$ and is give by $m_a$ to first order in  $v_a^2/M_{\tilde W}^2$.
The left-handed component of the charged lepton $a^- $  mixes  with the charged components of the adjoint triplet $ \psi_{\tilde W},$ 
that is:
\begin{equation}
 a'^{ -}= \cos{\phi}  \ a^{-}+ \sin{\phi} \  \psi^{-}_{\tilde W},
\end{equation}
where  the mixing angles are:
\begin{align}
 \cos{\phi}&= - \frac{ \sqrt{2} M_{\tilde W} } { \sqrt{(2 M_{\tilde W}^2+ g^2 v_{a}^2)}} \sim  -1 +g^2 \frac{ v_a^2}{M_{\tilde W}^2}+O( \frac{ v_a^2}{M_{\tilde W}^2}), \\
 \sin{\phi}&=   \frac{ g \ v_{a} } { \sqrt{( 2M_{\tilde W}^2+ g^2 v_{a}^2)}}  \sim  \frac{ v_a}{M_{\tilde W}} +O( \frac{ v_a^2}{M^2_{\tilde W}}).
 \label{eq:chargedmixing}
\end{align}
In the same way the neutrino $ \nu_a$ corresponds to the lightest neutralino.
 In the gauge-eigenstates basis $ \Psi^0_1=(\tilde B, \tilde W^0, \tilde H^{0}_d),$ and $  \Psi^0_{-1}=( \tilde H^0_u,  \nu_{a},  \psi^0_{\tilde B},  \psi^0_{\tilde W} )$ the neutralinos mass  term  has the form:
  \begin{equation}
  \mathcal{L}_N= -\frac{1}{2} (\Psi^0_{-1})^ T M_N \Psi^0_{1}+c.c.,
  \end{equation}
  where the mass matrix is:
 \begin{equation}
 \label{eq:neutmass}
M_N =
\begin{pmatrix}
 
   \frac{g' v_u }{ \sqrt{2}}   & - \frac{g v_u }{ \sqrt{2}}  & - \mu \\
 \frac{g' v_{a } }{ \sqrt{2}}   & - \frac{g v_{a } }{ \sqrt{2}}    & 0  \\
   M_{\tilde B}  & 0  &0 \\
 0 & M_{\tilde W}   & 0   \\
 \end{pmatrix}
\end{equation}
Then, the physical  neutrino corresponds  to the following mixture:
\begin{equation}
 \nu'_{a}= c_{\nu} \nu_{a}+ c_{ \tilde B} \psi_{\tilde B} + c_{\tilde W} \psi_{\tilde W},
 \end{equation}
 where the mixing angle:
 \begin{align}
 c_{\nu} &=-   \frac{1}{ \sqrt{ \frac{1}{2}\left(\frac{g v_a}{M_{\tilde W}}\right)^2+ \frac{1}{2} \left(\frac{g' v_a}{  M_{ \tilde B}} \right)^2+1}},    \\
 c_{ \tilde B}&= - \frac{g' v_a}{\sqrt{2}  M_{ \tilde B} \sqrt{ \frac{1}{2}\left(\frac{g v_a}{M_{\tilde W}}\right)^2+  \frac{1}{2}\left(\frac{g' v_a}{  M_{ \tilde B}} \right)^2+1 }} , \\
 c_{\tilde W}&=\frac{g v_a}{\sqrt{2} M_{\tilde W}\sqrt{\frac{1}{2}\left(\frac{g v_a}{M_{\tilde W}}\right)^2+  \frac{1}{2}\left(\frac{g' v_a}{  M_{ \tilde B}} \right)^2+1}} ,
  \label{eq:neutralmixing}
 \end{align}

 \section{ \label{constraints}Constraints from electroweak precision measurement}
 
  In the present section we will discuss constraints on our models from electroweak precision measurements (EWPM) and we will show that the MMRSSM parameter space compatible with the EWPM is large.
 First, we will present bounds on the sneutrino vev coming from lepton mixing and
subsequentely  we will discuss the EWPM limits on the down type Yukawa couplings that then translate in upper bounds on the sneutrino vev.
 
 As we showed in the previous section, the MMRSSM the  charged lepton $ a^-, $  and the neutrino $ \nu_{a} $   mix with the adjoint fermions as they both carry $R$ charge $\pm 1.$
The mixing changes the coupling of the lepton of flavour $a $  to the vector bosons and this will lead to deviations in predictions for EWPM. It is therefore essential to check under which conditions they are compatible with observations.
\par
The mixing of the charged lepton of flavour $a$ to the triplet leads the following modifications to its coupling to the  $Z$ boson:
\begin{equation}
 \mathcal{L}_{\text{NC}} = \frac{g}{2 \cos \theta_W} \bar{\psi}_a \gamma^\mu \left({g_V}_{\text{SM}}^a + \delta g_V^a- \left({g_A}_{\text{SM}}^a + \delta g_A^a \right) \gamma^5 \right) \psi_a Z_\mu
 \end{equation}
 where $\psi_a$ is the Dirac 4-component spinors for the charged lepton of flavour $a$, while the corrections to the Standard Model coupling 
  can beexpressed  in terms of the mixing angles of  equation \eqref{eq:chargedmixing}:
\begin{equation}
\delta g_V^a=\delta g_A^a=  - \frac{\sin^2{\phi}}{2} . 
 \label{eq:deltag}
\end{equation}
\begin{figure}
\centering
 \includegraphics[width=70mm]{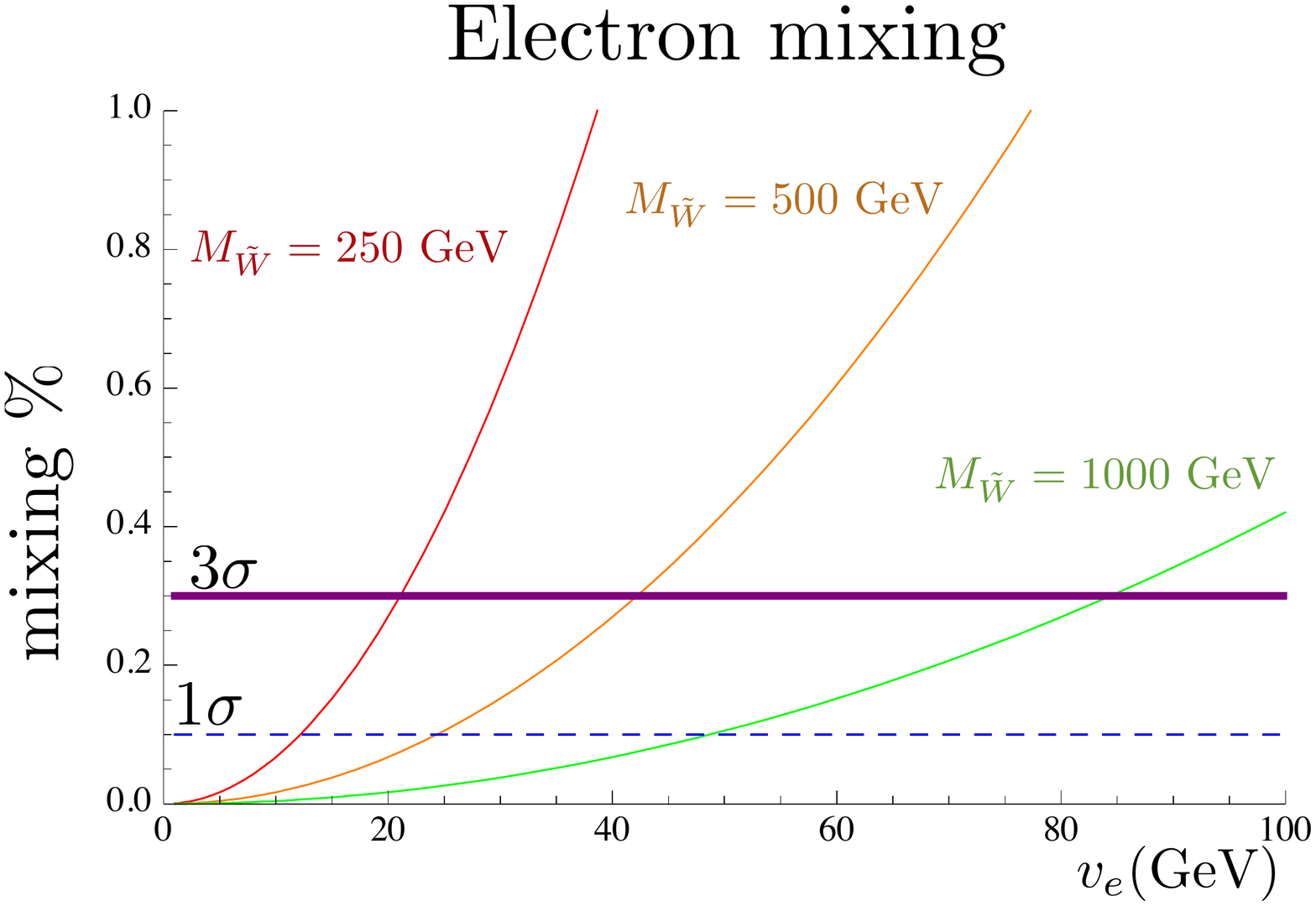}
\includegraphics[width=70mm]{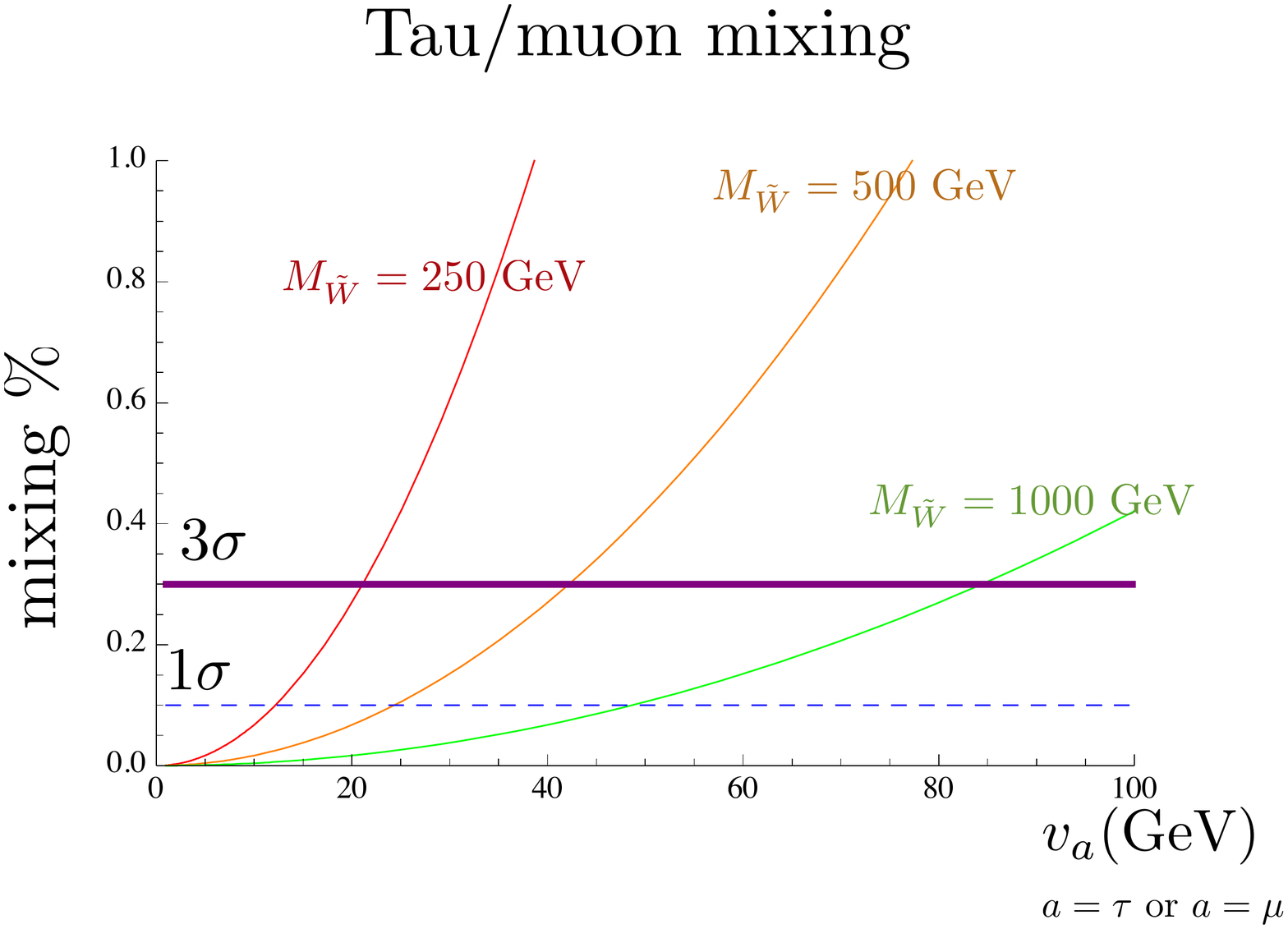}
\caption{ The lepton mixing with the triplet  (right electron and left for the muon or for the tau) taking into account different values 
of the wino mass  ($M_{\tilde W}= 250$ GeV (purple), $ M_{\tilde W}= 500$ GeV (orange), and $ M_{\tilde W}= 1000$ GeV (red)).
The blue horizontal line represents the $ 1 \sigma $ threshold, and the green one the $ 3 \sigma $ threshold.   }
 \label{fig:leptcoup}
\end{figure}

We can compare these corrections to the measured values of $g_V^a$ and $g_A^a$  \cite{:1970zzc} shown in table \ref{tab: vector-axial lepton couplings}. If  we  impose that  $ \delta g^a_V, $ and $ \delta g^a_A $ be within  the experimental error,
 we obtain that a mixing smaller than $ 0.07 \% $ is tolerated at $ 1 \sigma$ level by EWPM when  $ a=e.$ For $ a=\mu,$ and $a=\tau$ ,
the limit is $ 0.1  \%.$ 
Inserting eq.(\ref{eq:chargedmixing}) in  eq.(\ref{eq:deltag}) we obtain bounds on the sneutrino VEV which are shown in fig.  \ref{fig:leptcoup}.  For winos at the electroweak scale the region allowed by the experimental data is a fairly high $ \tan{\beta} $ region  $( \tan{\beta}> 11)  $  at  $1 \sigma $ level.
 However, it is possible to enlarge the  parameter space by considering heavier gauginos, for example $  M_{ \tilde W} = 1 $ TeV $ requires only  \tan{\beta}> 2. $  Therefore, the MMRSSM tends to favor a scenario with fairly heavy gauginos .
 \begin{table}[t]
\center
\label{tab: vector-axial lepton couplings}
\begin{tabular}{|c||c||c|}
\hline
Lepton & $ g^l_V $ & $ g^l_A $ \\
\hline
$ e $ & $ -0.03817 \pm 0.00047$ & $ -0.50111 \pm 0.00035$  \\
$ \mu$ & $ -0.0367 \pm 0.0023$ & $ -0.50120\pm 0.00054$\\
$ \tau $ & $-0.0366 \pm 0.0010$ & $ -0.50204\pm 0.00064$\\
\hline
\end{tabular}
\caption{Effective vector-axial lepton couplings.}
\end{table}

\par
\begin{figure}[t]
\begin{center}
\includegraphics[scale=0.35]
{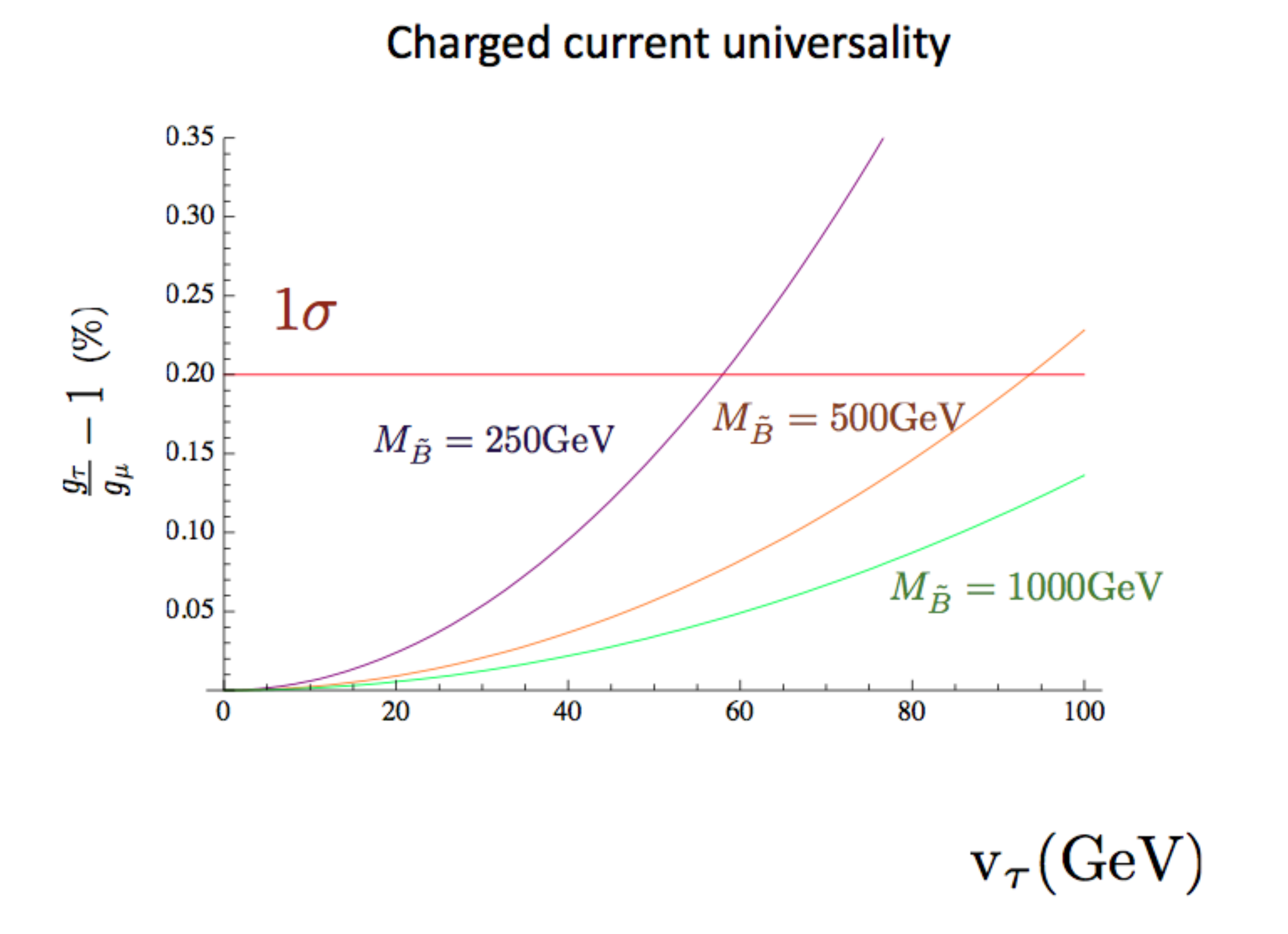}
\end{center}
\caption{  
The violation of leptonic universality in the charged current interaction  $  1- \frac{ g_{\tau}}{g_{\mu}} $ assuming the the mixed lepton is the $ \tau.$
We considered values of $ M_{\tilde W}= 250$ GeV (purple), $ M_{\tilde W}= 500$ GeV (orange), and $ M_{\tilde W}= 1000$ GeV (red).
The blue horizontal line represents the $  0.3 \% $ threshold, and the green one the $ 3 \sigma $ threshold.  }
  \label{fig:chargeduniv}
\end{figure}

Since only one of the flavour mixes with the triplet, lepton universality is broken in our model. Charged current universality is verified experimentally  to
 the $ 0.2  \% $  level for both $ e-\mu $, and  $\mu-\tau$ \cite{Pich:1997hj,Loinaz:2004qc}, but we find that we do not obtain stronger bounds from this fact  than those derived from the $Z$ coupling. This is shown in fig  \ref{fig:chargeduniv} where we  plotted, taking $a= \tau,$  $  \frac{g_{\tau}}{ g_{\mu}} -1 $  where:
\begin{align}
  \frac{g_{\tau}}{ g_{\mu}} = \cos{\phi} \ c_{\nu}+ \sqrt{2}  \sin{\phi}\ c_{\psi_{\tilde W}}.
   \label{eq:chargeduniv}
  \end{align}
 
In the MMRSSM the down-type Yukawa couplings  give extra tree level contributions to  electroweak observables which put constraints one those couplings, and  therefore  put  a lower  bound on the sneutrino vev.
As we have already noticed in the previous section, the MMRSSM down-type Yukawa couplings have the same form as standard $ R_p $ violating trilinear couplings. Indeed, the lepton Yukawa couplings  correspond  in the standard notation to the 
$ \lambda_{ijk} L_i L_j E^c_k $ couplings, while the down type quark   Yukawa couplings correspond to $ \lambda'_{ijk} L_i Q_j D^c_k .$ 
Therefore, these extra tree level contribution to the electroweak observables are the same as in  standard $ R_p $ violating models and we can use result from the literature on those models (see \cite{Barbier:2004ez} for a review) to put bounds on the Yukawa couplings of our model.
 \par
The strongest bound  when $ a= e $ or $ a=\mu $ comes from the tau Yukawa coupling $ L_e L_{\mu} E^c_{\mu} $( or $ L_{\tau} L_{\mu} E^c_{\mu} $)
These operators lead to an additional contribution to the leptonic tau decays via $ \tilde \tau_R $ exchange.  This affects the ratio $ R_{\tau \mu},$ defined as:
 \begin{equation}
 R_{\tau \mu}= \frac{ \Gamma( \tau \rightarrow \mu \nu \nu)}{ \Gamma( \tau \rightarrow e \nu \nu)},
 \end{equation}
 and  leads to the following bound:
 \begin{align}
 y_{\tau} < 0.07 \left(\frac{100 \text{GeV}}{m_{\tilde \tau^c}}\right)^2.
 \end{align}
for   $ m_{\tilde \tau^c}= 100 $ GeV. This bound implies a lower limit for the sneutrino vev  $  v_{a}>15$ GeV  both for $a=e,$ and for $ a=\mu.$ 
We see that this would  exclude the region of the parameter space with gauginos with a
mass around the electroweak scale. Therefore,  the MMRSSM spectrum is characterised by fairly heavy gauginos or in another words the very high $ \tan{\beta} $ region in the MMRSS is excluded by the experimental constraints 
on the Yukawa coupling.
\par
When $ a=\tau$ the strongest bound on the sneutrino vev comes from the bottom Yukawa coupling.
The trilinear coupling $ L_a Q b^c $ leads to an additional contribution to at loop level to the partial width of the $Z$ to $\tau$.
The comparison with  experiment gives the following bound:
\begin{equation}
|y_b| <  0.58 \left( \frac{m_{\tilde b_R}}{100 \text{GeV}}\right)^2 \; \;.
\end{equation}
Therefore, the MMRSSM parameter space for $ a= \tau $ is less constrained, and in particular it contains also a very $\tan{\beta} $ region.
\par
 In the standard $R_p$ violating scenario,  the EWPM bounds are subleading compared
to the bounds that come from the generation of Majorana mass for neutrinos.
If we consider for example   $ \lambda'_{a33} = y^a_b , $ that is the bottom Yukawa coupling in our model,
we see that the constraints on the neutrino mass require: $ \lambda'_{a33} > 10^{-6},$ while in our case the same coupling  can be several orders of magnitude bigger: $  y^{\tau}_b >0.58. $ 
We will investigate the phenomenological consequences of this in s ection 5.
\par
Standard  $R_p$ violating trilinear couplings are also constrained by  cosmological bounds and these constraints can be quite stringent.
For example, the requirement that an existing baryon asymmetry is not erased before the electroweak transition typically implies \cite{Dreiner:1997uz}
$ \lambda, \lambda' < 10^{-7}.$ These constraints do not apply to our case, as the model preserves the baryonic number as well as  lepton number. However,
as we will see in the following section, the MMRSSM requires a very low re-heating temperature and would require a different baryogenesis mechanism.

 \section{\label{Rbreaking} $R$-symmetry breaking }
  $R$-symmetry is not an exact symmetry because it is broken (at least) by the gravitino mass  term that is necessary to cancel the cosmological constant. This breaking is then communicated to the visible
sector, through  anomaly mediation if nothing else. 
Therefore, we need to take into account  the following additional anomaly-mediated,  $R$-symmetry violating soft terms \cite{Martin:1997ns}:
\begin{align}
\label{eq:amsb}
\mathcal{L}_{AM}&= A^u \tilde u_
r \tilde q_L H_u-A^d \tilde d_R \tilde q_L \tilde l_a
-A^l \tilde l_a \tilde l  \tilde  e_R +  \\ \nonumber
& M_{\lambda_{\tilde B}} \lambda_{\tilde B} \lambda_{\tilde B}+ M_{\lambda_{\tilde W}} \lambda_{\tilde W} \lambda_{\tilde W}+ M_{\lambda_{\tilde g}} \lambda_{\tilde g} \lambda_{\tilde g},
\label{eq:AM}
\end{align}
where:
  \begin{align}
  M_{\lambda_{i}}= \beta_i  \frac{\alpha_i}{ 4 \pi } m_{\frac{3}{2}},
   \end{align}                                                                                                                                                                              
   \begin{align}
   A_{ijk}=- \beta_{y_{ijk}} m_{\frac{3}{2}},
   \end{align}
   where $  m_{\frac{3}{2}} =\frac{\Lambda^2}{  M_{P} } $ is the gravitino mass, and $ \Lambda \sim \sqrt{D'}$ indicates the SUSY breaking scale.
   Therefore, the gauginos are not pure Dirac fermions, but pseudo Dirac.
For relatively low SUSY breaking  scale $ \Lambda $    these contributions  will be subdominant compared to the $R$-symmetric SUSY breaking terms  in equation \eqref{eq:softsusy} and will not have important phenomenological consequences. One important exception is that they will generate neutrino masses that can be above the present bound. Also, the presence of a massive gravitino which in our case is unstable leads to important bound on the reheating temperature.
\subsection{\label{neutrino}Neutrino masses}
The SUSY breaking term of equation \eqref{eq:amsb} also break the $U(1)_R$ symmetry and will inevitably generate a Majorana mass term for the neutrino of flavour $a$, and this will translate to a limit on the SUSY breaking scale.    
  \par
  \begin{figure} 
\centering
 \includegraphics[width=70mm]{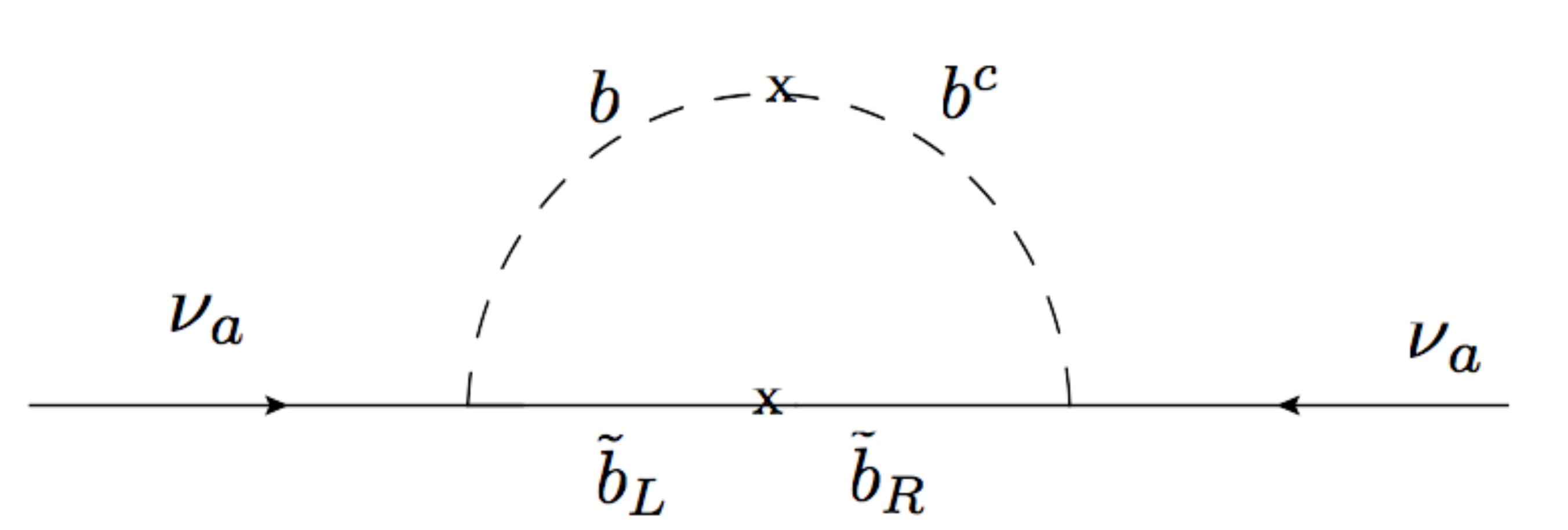}
\includegraphics[width=70mm]{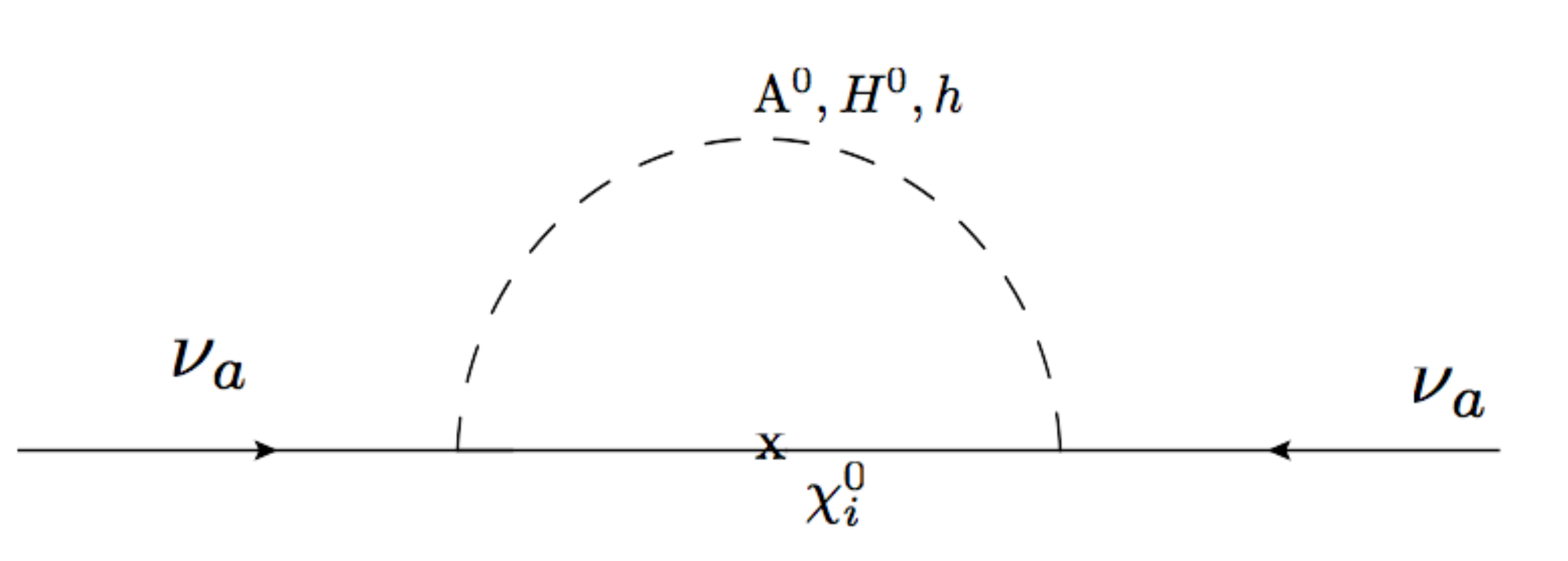}
\caption{ Majorana mass term for the neutrino $ \nu_a $ generated at one loop from anomaly mediation ( by gaugino Majorana mass on the right, and left/right mixing on the left). }
\label{fig:neutrinomass}
\end{figure}

   At tree level the neutrino remains massless. Indeed, even after introducing the Majorana masses $M_{\lambda_i}$ for the gauginos in the neutralino mass matrix of equation \eqref{eq:neutmass} the smallest eigenvalue is still zero.
At one loop a Majorana mass term for $ \nu_a$ is induced by the diagrams in fig.\ref{fig:neutrinomass}. The contribution coming from the insertion of an $A$ term is given parametrically by (see \cite{Barbier:2004ez} for the full expression):
  \begin{equation}
    M_{\nu_a} \sim 3 \left(\frac{1}{16 \pi^2} \right)^2 \left( \frac{m_b}{m_{\tilde{b}}}\right)^2 y_b^2 m_{\frac{3}{2}} 
    \label{eq: neu1}
  \end{equation}
     where $ \tilde m_b $ is an averaged sbottom mass parameter.
 The mass contributions in equation \eqref{eq: neu1} is suppressed by the Yukawa couplings that assume their maximum values at large $\tan \beta$. For example, when $ v_a \sim $5 GeV and $m_{\tilde b}  \sim 200$ GeV,
 requiring $M_{\nu_a} \lesssim 1$ eV leads to: $  m_{\frac{3}{2}}  \lesssim 10 $  MeV which implies $ \Lambda \lesssim 10^{8} $ GeV.
  The contribution from the diagram with a Majorana gaugino mass insertion is given parametrically in the large $\tan \beta$ limit by:
 \begin{equation}
 M_{\nu_a} \sim \left(\frac{1}{16 \pi^2} \right) \frac{m_Z^2}{m_{\chi_0}^2} \frac{M_{\lambda}}{\tan^2 \beta}
 \end{equation}
 where $m_{\chi_0}$ is the neutralino mass and $M_{\lambda} \sim m_{\frac{3}{2}}/(16 \pi^2)$ is the Majorana gaugino mass insertion. The corresponding bound is then stronger for lower $\tan \beta$. For $v_a = 100$ GeV and $m_{\chi_0} = 1$ TeV, 
 asking for $M_{\nu_a} \lesssim 1$ eV leads to $m_{3/2} \lesssim 10$ MeV which implies $\Lambda \lesssim 3 \times 10^{7}$ GeV.Therefore, the MMRSSM is compatible with the bounds on the neutrino masses, as long as we consider a fairly low SUSY breaking scale like  a scenario of gauge mediated SUSY breaking. 
  
Neutrino masses for the other flavour $b,c$ can be introduced through higher dimensional operators of the form:
\begin{equation}
\int{ d^2\theta \frac{ ( H_u L_{b,c})( H_u L_{b,c})}{ M_f}}, \\ 
\end{equation}
where the scale $ M_f $ is a flavor scale where the overall lepton number $ L_b + L_c$ is broken. 
\subsection{Cosmological bounds: gravitino LSP}
\begin{figure}
\begin{center}
\includegraphics[scale=0.4]
{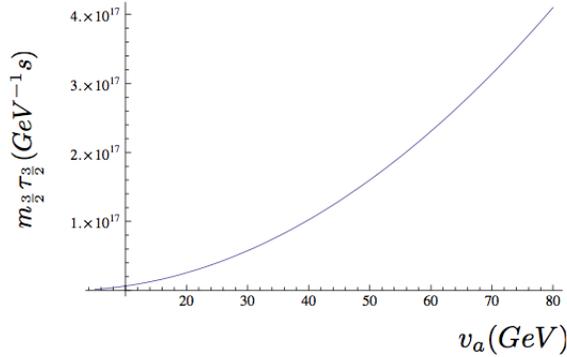}
\end{center}
\caption{Gravitino mass times its lifetime. }
  \label{fig:MT}
\end{figure}

In the MMRSSM the gravitino is the lightest supersymmetric particle. Indeed the bounds on the neutrino mass constrain it to be lighter than $ \sim 1$ MeV.
In our model the gravitino  is unstable  and decays to a neutrino of flavor $a$ and  a monochromatic photon.
Therefore, it is necessary to evaluate its cosmological impact. This requires first computing the gravitino life time.
The tree level  contribution for the decay $ \tilde G \rightarrow  \gamma \nu_a $  \cite{Takayama:2000uz} is given by :
\begin{align}
\Gamma_{tree} ( \tilde G \rightarrow  \gamma \nu_a ) \sim  \frac{ | U_{\tilde B \nu_a} |^2 m^3_{3/2}}{ 32 \pi M_P^2}
\end{align}
where $ U_{\tilde B \nu_a} $ is the mixing between the neutrino and the bino and is proportional to the neutrino mass.
 The leading contribution comes instead from a one loop diagram  and is given by \cite{Borgani:1996ag}:
\begin{align}
\Gamma( \tilde G \rightarrow  \gamma \nu_a )= \frac{ \alpha \  y_b^2}{128 \pi^4}   \frac{ m_b^2  m_{\frac{3}{2}} }{M^2_{P}}  \ln^2{\frac{ m_{\tilde b}}{m_b}}=
 \frac{ \alpha \  v^2_a}{128 \pi^4}   \frac{   m_{\frac{3}{2}}  m^4_b}{M^2_{P}}  \log^2{\frac{ m_{\tilde b}}{m_b}} \;.
 \label{eq:width32}
\end{align}
The gravitino  lifetime increases with the sneutrino vev $ v_a, $ and decreases with the gravitino mass .The 
lifetime of a  $ 1$ MeV is approximately  $ 10^{20} $  s for a sneutrino vev of $40$ GeV (see figure \ref{fig:MT}) . So, the gravitino 
 lifetime is larger than the lifetime of the universe $\sim 10^{17}$s,  and this means that it could be a  dark matter candidate.
However, because it is unstable, its abundance is constrained by the observed  $\gamma,$ and x rays background and $\gamma$ ray lines from the milky way.
\par
Searches for gamma-ray lines from the galactic center \cite{Yuksel:2007dr} put a model independent bound on the mass times the lifetime of an unstable dark matter particle decaying to a monochromatic photon. For photon energy between $\sim 10^{-2}$ MeV and $\sim 10$ MeV, the bound is approximately $m_{3/2} \tau_{3/2} \sim 10^{28}$ GeV s. So, for a gravitino mass consistent with the neutrino mass bound: $m_{3/2} \sim 1$ MeV, the bound is $\tau > 10^{28} s$, well above what the estimate that equation \eqref{eq:width32} indicates. This means that the gravitino cannot on its own be the dark matter. We can obtain a bound on the gravitino abundance by rescaling the bound of \cite{Yuksel:2007dr}  and making use of \eqref{eq:width32} for the lifetime \footnote{The bound of  \cite{Yuksel:2007dr}  depends on the actual dark matter halo, so a simple rescaling is only approximate}:
\begin{equation}
\Omega_{3/2} h^2 <  7 \times 10^{-11} \left(\frac{v_a}{1 \text{GeV}} \right)^2 \frac{1}{ \log^2(m_{\tilde b}/m_b)}
\label{eq:Omegaboundline}
\end{equation}
which is independent of the mass and which for $v_a = 40$ GeV gives $\Omega_{3/2} h^2 \lesssim 4 \times 10^{-9}$.

The bound from the diffuse photon background is weaker. For a gravitino with dark matter abundance: $\Omega_{3/2} h^2 \sim 0.1$, the bound on $m_{3/2} \tau_{3/2}$ is $\sim 10^{23}$ GeV s \cite{Yuksel:2007dr} ,which , since the photon flux from dark matter decay is proportional to $\Omega_{3/2}/(m_{3/2} \tau_{3/2})$, can be turned into a bound on $\Omega_{3/2}$:
\begin{equation}
\Omega_{\frac{3}{2}} h^2 <  7 \times 10^{-9} \left(\frac{v_a}{1 \text{GeV}} \right)^2 \frac{1}{ \log^2(m_{\tilde b}/m_b)} 
\end{equation}
for $m_{\frac{3}{2}} \sim 1$ MeV. Notice however that unlike the gamma-ray line bound, the diffuse photon bound depends on the energy of the emitted photon $\sim m_{\frac{3}{2}}/2$. For example, for $100$ keV gravitino, the bound on $\Omega_{\frac{3}{2}}$ is reduced by almost two orders of magnitude.
%
\par
\begin{figure}[h]
\begin{center}
\includegraphics[scale=0.4]
{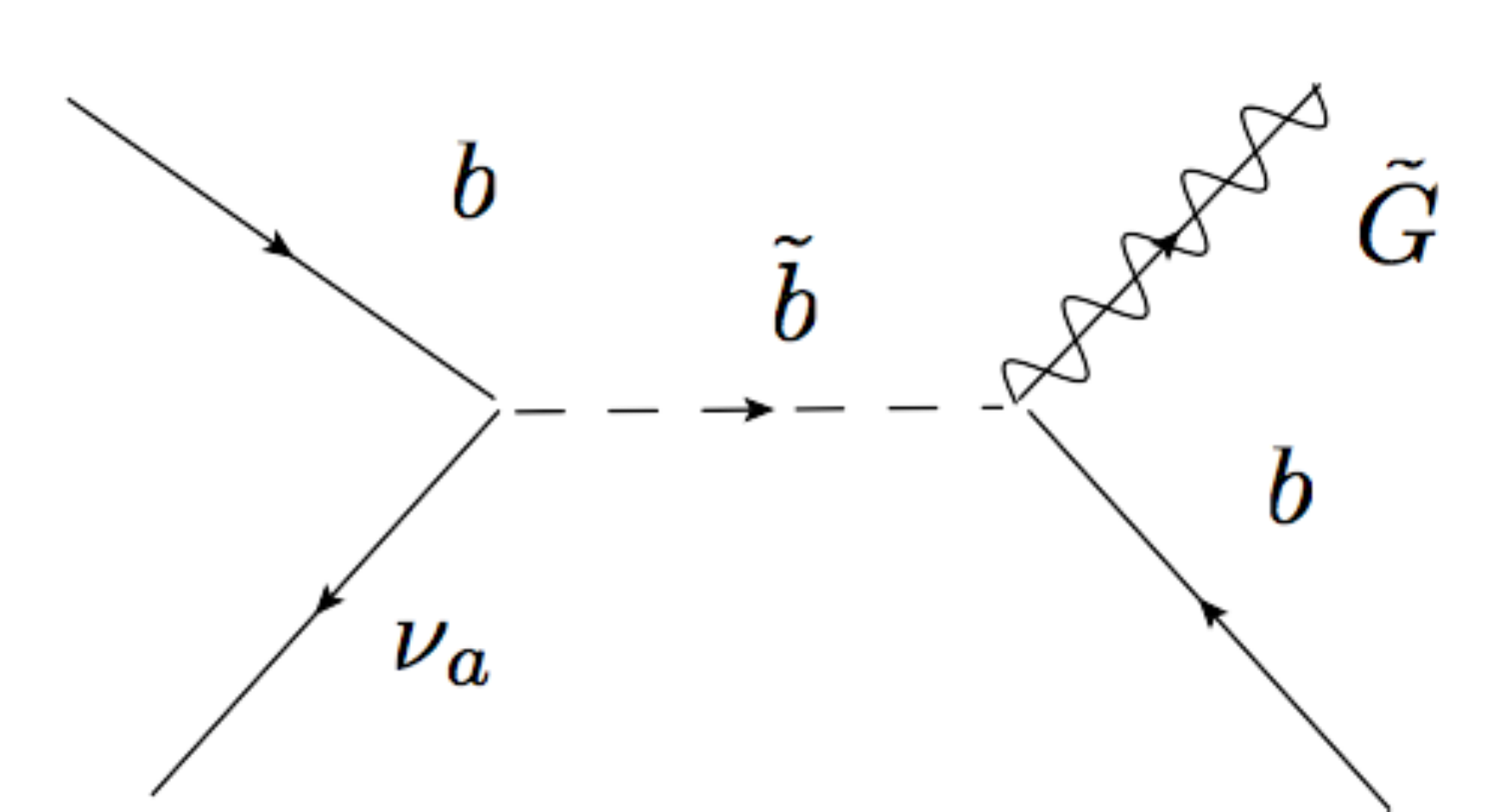}
\end{center}
\caption{Gravitino scattering process that will generate the gravitino abundance after reheating}
  \label{fig:gravprod}
\end{figure}

To respect those strong bounds on the gravitino relic density, one must assume a low reheating temperature $T_{\text{RH}}$. If it is above the SUSY scale, gravitino will be produced through scattering with the thermal plasma which includes superpartners, and it's relic density will be given by \cite{Giudice:2000ex}:
\begin{align}
\Omega_{\frac{3}{2}} h^2= 0.13 \left(  \frac{T_{\text{RH}}}{10^{5} \text{GeV}}\right) \left(  \frac{1 \text{MeV}}{m_{3/2}} \right) \left( \frac{m_{\tilde g}}{1 \text{TeV}} \right)^2,
\end{align}
where $ m_{\tilde g} $ is the gluino mass.
This yields a relic abundance that is too large to satisfy our constraint.
Therefore, we need to 
consider a scenario with a reheat temperature  that  is below the SUSY threshold.
In this scenario the gravitinos are produced by the thermal scattering of neutrino and bottom quark (see figure \ref{fig:gravprod}) with a cross section given parametrically by:
\begin{equation}
  \sigma \sim y_ b^4 \frac{T_{\text{RH}}^6}{\Lambda^4 m_{\tilde b}^4},
\end{equation}
  and the relic density is given by \cite{Giudice:2000ex}:
\begin{align}
\Omega_{\frac{3}{2}} \sim 10^{24}  y_b^2    \frac{m_{3/2} T^7_\text{RH}}{\Lambda^4 m_{\tilde b}^4},
\end{align}
 and when combine with \eqref{eq:Omegaboundline}  this yields the following bound on the reheat temperature:
\begin{equation}
T_\text{RH} \lesssim  70 \text{GeV} \left(\frac{m_{3/2}}{1 \text{MeV}}\right)^{1/7} \left(\frac{m_{\tilde b}}{200 \text{GeV}} \right)^{4/7} \left(\frac{v_a}{30 \text{GeV}} \right)^{6/7} \;.
\end{equation}
Because of the required low reheating temperature, 'standard' baryogenesis or leptogenesis scenario will not work. One might be able to accomodate a larger reheat temperature if the gravitino decays to a hidden sector instead of a photon and a neutrino, or if the gravitino density is somehow diluted at a late time.
\section{\label{Rsusybreaking}$R$-symmetric  gauge mediation}
The bounds on the neutrino masses from section \ref{neutrino} require a low  SUSY breaking scale. This means that high scale SUSY breaking mechanism such as gravity or anomaly mediated SUSY breaking will not work in our context and R-symmetric gauge mediation is a more natural possibility. R-symmetric gauge mediation was studied in some details in \cite{Benakli:2008pg}
 and \cite{Benakli:2009mk}. One of the main issue is to generate positive masses for the adjoint scalar. This can be achieved with appropriate choice of couplings between the adjoint superfields and the messengers. In the MMRSSM, the $\mu/B_{\mu}$ problem takes a slightly different form, and the susy breaking mediation mechanism also need to generate the Yukawa coupling for the lepton of flavour $a$ which is not generated in the low energy theory. 

\subsection{ $ R$ symmetry, and the  $ \mu /  B_{\mu} $ problem}
In $ R$ symmetric models  the $ \mu /  \ B_{\mu} $ problem is a different problem than in the MSSM\footnote{ For a discussion of the $\mu B_{\mu}$ problem in model with Dirac
gaugino, but with $R$ symmetry breaking in the Higgs sector see \cite{Benakli:2011kz}}  . Indeed,  the $ \mu, $ and the $ B_{\mu} $ terms contain different fields and therefore, they can be generated by separate
UV physics.
 For example, in the MMRSSM the $\mu $ term is $ \mu \tilde{H_u} \tilde R_d, $ while the   $ B_{\mu} $ term is  $ B_{\mu} H_u \tilde l_a.$  In the MRSSM, instead,
 the  $ \mu $ term has the form $ \tilde R_u \tilde H_d+\tilde H_u \tilde R_d,$ while  the   $ B_{\mu} $ term is  $ B_{\mu} H_u H_d$ \cite{Amigo:2008rc} (see also \cite{Nelson:2002ca} for a model without a $\mu$ term).
This facilitate the generation of  the $ \mu $ term at one loop, and the $ B_{\mu} $ term at two loops. However, as we will see below, this is not sufficient to assure the naturalness of the model.
\par
If we assume that SUSY is broken  only  by the $ D$-term of vector superfield spurion, the effective operators which generate the  $ \mu $ and $  \ B_{\mu} $  terms are:
\begin{align}
&\frac{1}{M^3}  \int{ d^4\theta (W^{' \alpha} W'_{\alpha} )^{\dagger} H_u R_d}, \\
&\frac{1}{M^6} \int{ d^4\theta (W^{' \alpha} W'_{\alpha} )^2H_u L_{a} }.
\label{eq:Bmubmuop1}
\end{align}
If $ M$ is the messenger mass scale, $ D' \ll M^2$ and  the $ \mu $ term is  too small: $ \mu \sim \frac{ D^2 }  { 16 \pi^2 M^3},$  unless $ D \sim M^2$ or $ D \sim 10^{-1} M^2 $ with the gauginos at the TeV scale. Another possibility is that the denominators of the operator \eqref{eq:Bmubmuop1} are made out of different mass scales, similar to the model of \cite{Dvali:1996cu}
We can for example write a superpotential of the form :
\begin{align}
W^D_{\mu B\mu}&=  M_{\Phi}   \Phi_{+} \Phi _{-} + M_S \bar S S +  M_N \bar N N+ \\  \nonumber
&+ S( \lambda_1 R_d H_u + \lambda  \Phi_{+} \Phi _{-} ) +  \lambda_2 S^2   N + \tilde \lambda_1H_u  L_{a}   N,
\label{eq:musuperpotentialD}
\end{align}
where  $  \Phi_{+,-} $ are messenger fields that are singlet under the SM gauge groups, which carry $ U(1)'$ charge,  R-charge $1$ and  get soft mass terms from the $D'$-term.
\begin{table}
\centering 
{%
 \begin{tabular}{cc} 
\multicolumn{1}{c|}{\textbf{SuperField}} & 
\multicolumn{1}{c}{\textbf{$U(1)_R$}} \\
\hline 
$ \Phi $ & 0  \\ 
$ \bar \Phi$  & 2  \\ 
$ S $& 0  \\ 
$ \bar S$  & 2  \\ 
$ \bar N $  & 0  \\ 
$ N $  & 2  \\ 
\end{tabular} 
}\qquad\qquad
\caption{R-charge assignement} 
\label{table:Rcharges}

\end{table} 

The other fields, $S,\bar S ,N,\bar N $ are all singlet under the SM gauge groups and have the $R$-charge assginment shown in table \ref{table:Rcharges}:
\par
The $ \mu $ term is then:
\begin{align}
\mu \sim \frac{\lambda \lambda_1}{16 \pi^2} \frac{{D'}^2}{M_T M^2_S}, 
\end{align}
where $M_T$ is the mass of the messenger scalars $\Phi_{+,-}$. If one then assumes $M_S \sim \sqrt{D'} ,$  on can get $\mu$ term at the weak scale or little bit above. The $B_\mu$ term needs to involve the superfield $N$ and will be generated at two loops with the same size as $\mu^2$.  In models with a SUSY breaking spurion with an $ F$-term: $X = \theta^2 F$,  the $\mu $ and the $ B_{\mu}$ terms could be generated through the following effective operators:
\begin{eqnarray}
&\frac{1}{M}  \int{ d^4\theta X^{\dagger} H_u R_d}, \\
\label{eq:muop}
&\frac{1}{M^2} \int{ d^4\theta (X^{\dagger} X) H_u L_{a} }.
\label{eq:Bmuop}
\end{eqnarray}
As usual, in order to avoid fine tuning problems, the $ \mu, $ and the $ B_{\mu} $ terms should be of the same order, that is $  B_{\mu}\sim \mu^2$ This means that $ B_{\mu} $ needs to be
generated at two loops, while $ \mu $ has to be generated  at one loop order.
\par
In our model the $ \mu$ and the $B_{\mu} $ terms are generated by operators different fields and this makes it easier to write down a superpotential, 
 that  possesses 
 accidental symmetries which  forbids the $ B_{\mu} $ term at one loop and allow, instead, the generation of the $ \mu$ term. But this is not sufficient to guarantee the naturalness of our model as
 the new couplings of the Higgs with the messenger sector can generate  additional soft mass terms $ m^2_{H_u}$ at one loop, which would be larger the one coming from the  $\mu$ term  by the square root of a one loop factor.
This problem can be addressed by considering a model analogous  to the one considered in \cite{Dvali:1996cu}  that do not couple the Higgs superfield directly to the messenger, but use some extra singlet to generate the $\mu$ term. In this case the $\mu$ term arises from an operator of the form:
 \begin{align}
&\int{ d^4\theta  \frac{ D^2( X^{\dagger} X)}{ M^3} H_u R_d } \; \;,
\label{eq:Giudicemu}
\end{align}
instead of \eqref{eq:muop}. The $B_\mu$ term also receives contribution from an operator of the form
\begin{align}
&\int{ d^4\theta  \frac{ (X^{\dagger} X) \bar D^2D^2( X^{\dagger} X)}{ M^6} H_u L_a }.
\label{eq:GiudiceBmu}
\end{align}
\par
Notice the scaling of those operators is very similar to the one in \eqref{eq:Bmubmuop1}.  They can also be generated through a very similar superpotential with one vector-like messenger field $\Phi,\bar \Phi$ and two singlet $N, \bar N$ and $S,\bar S$: 
\begin{align}
W_{\mu B\mu}&=  M_{\Phi}  \bar \Phi \Phi+  M_S \bar S S +  M_N \bar N N+   \tilde  \lambda X \Phi \Phi+ \\  \nonumber
&+ S( \lambda_1 R_d H_u + \lambda  \Phi  \bar \Phi) +  \lambda_2 S^2   N + \tilde \lambda_1H_u  L_{a}   N,
\label{eq:musuperpotential}
\end{align}
where $ M_{\Phi} \sim M_T$ is  the messenger mass scale and $ M_S \sim M_N \sim \sqrt{F}.$ The $R$-charge assignment is again the same as the one shown in table \ref{table:Rcharges}. The superpotential \eqref{eq:musuperpotential} will not generate an operator of the form \eqref{eq:muop} since it has a $U(1)$ symmetry under which $\Phi$ and $\bar \Phi$ have charge $\pm 1$ while $X$ as charge $-2$. One can also easily show by examining the various spurious $U(1)$ of the superpotential that the $B_\mu$ term can only arise at two loops.
\begin{figure}[h]
\begin{center}
\includegraphics[scale=0.4]
{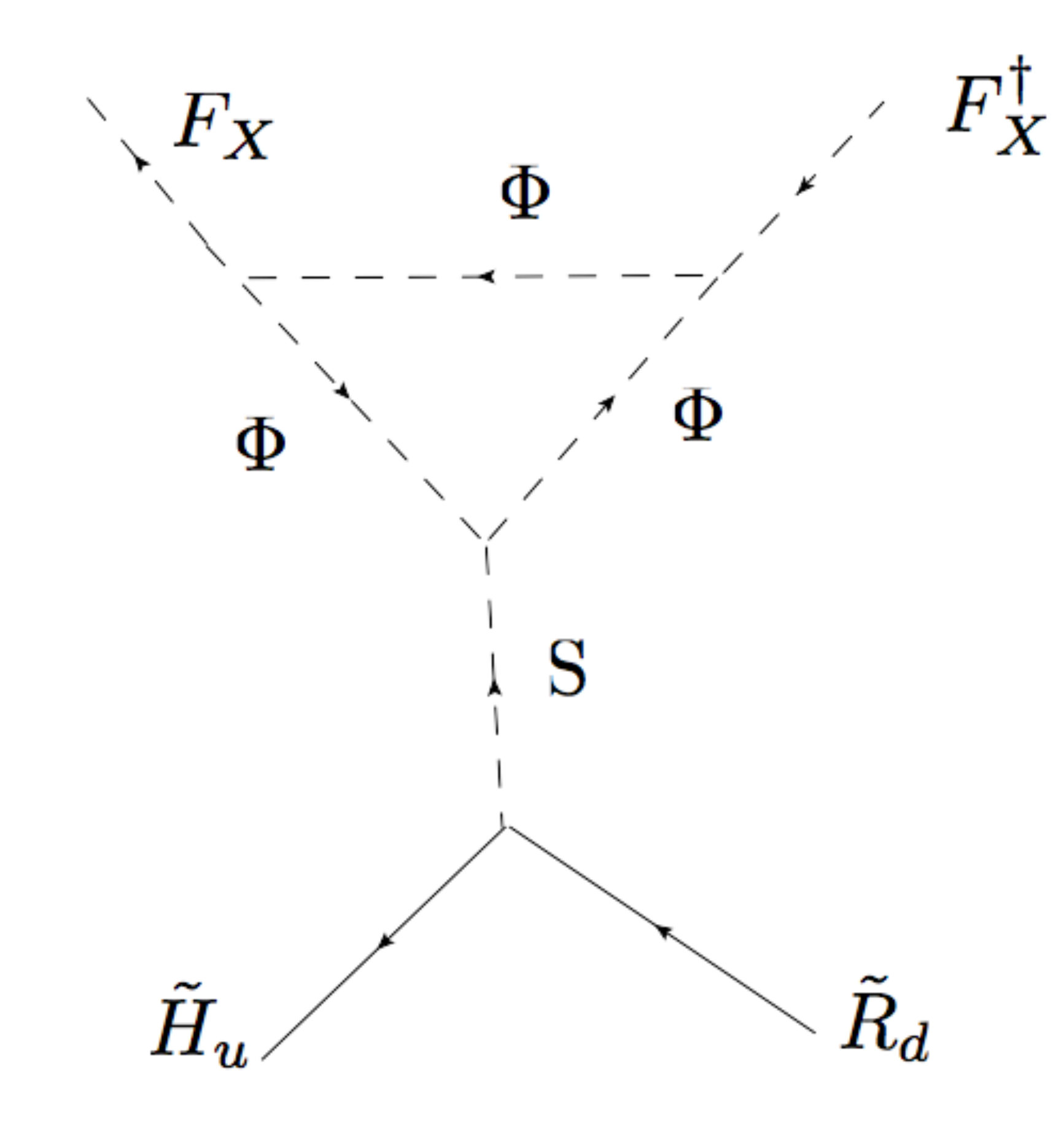}
 \includegraphics[scale=0.4]
{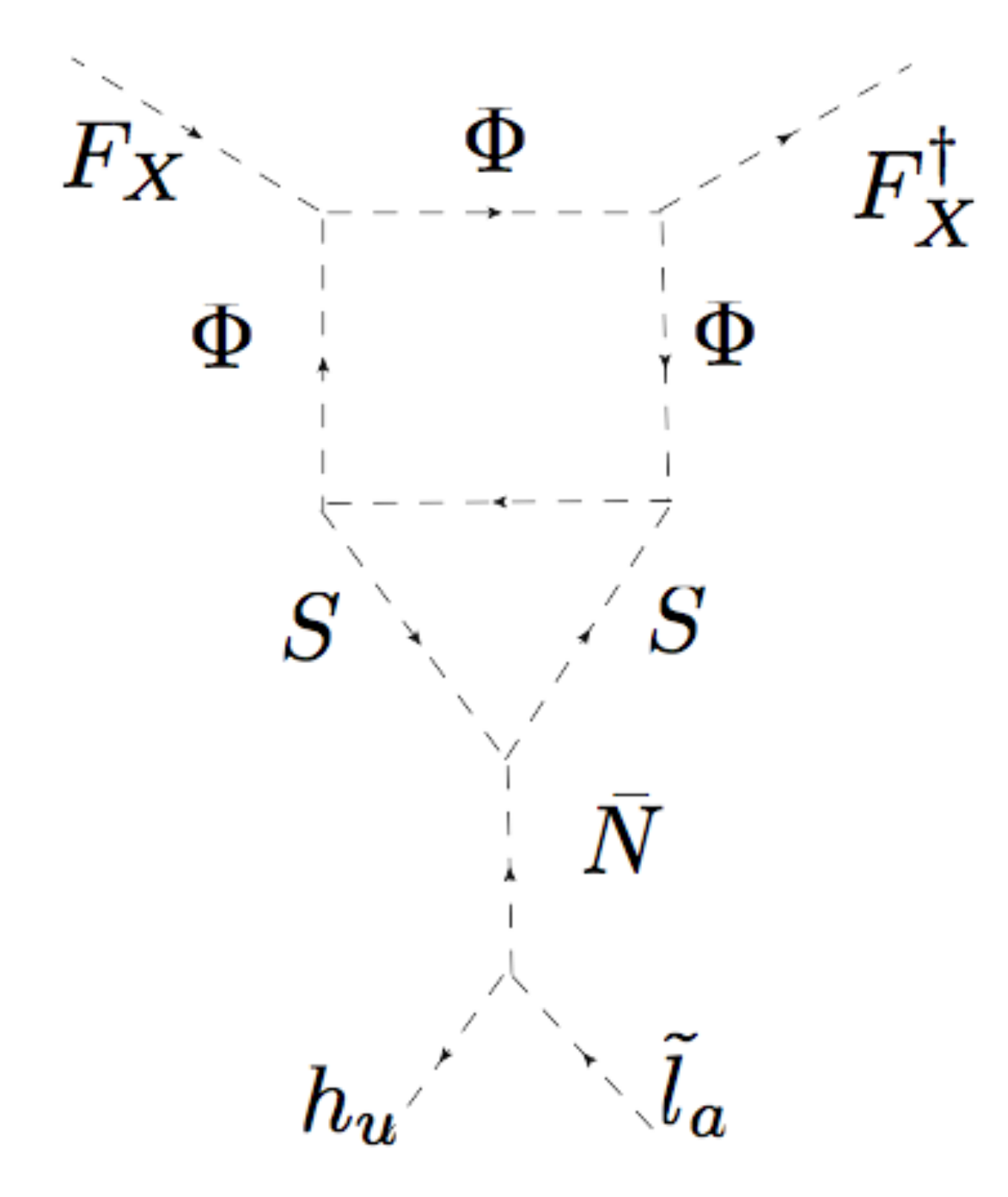}
\end{center}
    \caption{ Contribution to the $ \mu$ and $ B_{\mu} $ terms generated by  \eqref{eq:musuperpotential}. }
    \label{fig:muterm}
\end{figure}

  The $ \mu $ term on the other hand  can be   generated from the one loop diagram in fig.\ref{fig:muterm} , and it is given parametrically by:
\begin{align} 
\mu \sim \frac{\lambda \lambda_1 {\tilde \lambda}^2}{16 \pi^2} \frac{F^2}{M_T M^2_S} \sim \frac{1}{16 \pi^2} \frac{F}{ M_T}
, \end{align}
withs $ M_S \sim \sqrt{F}. $ The $ B_{\mu} $ term can instead be generated by the two loops  diagram in figure \ref{fig:muterm}, :
\begin{align}
B_{\mu} \sim \frac{\tilde \lambda_1 \lambda^2  \lambda_2}{ (16 \pi^2)^2} \frac{ F^2}{ M^2_T} \sim \mu^2.
\end{align}
Summarizing, this mechanism allows us to generate the $ \mu $ term at one loop, and the $ B_{\mu} $ together with the scalar masses to be generated at two loops. In order to avoid fine tuning problems we have to introduce a third
scale
$ M_S^2 \sim F, $  and several link fields. However, we will see in the section below that these link fields are important also to generate the Yukawa couplings.
\par
To avoid the introduction of the extra link fields, we would need to consider a model $m_{H_u}^2 \gg \mu^2,B_{\mu}$, which in an otherwise completely natural model would require some fine-tuning to achieve the correct pattern of electroweak symmetry breaking. However, since, as we have mentioned previously, we already seem to require fine-tuning to evade the LEP Higgs bound, this hierarchy might in fact not introduce an extra source of fine-tuning (see \cite{DeSimone:2011va} for a related idea).  

\subsection{ Yukawa coupling for lepton $a$}
As we have already explained, the Yukawa coupling for the lepton of flavor $a$ needs to be generated by the SUSY breaking sector.
In models with an $F$-term SUSY breaking spurion, it can be generated by the following operator:
\begin{align}
\int{\frac{d^4 \theta}{M^2} X^{\dagger} H_u^{\dagger} L_a l^c_a},
\label{eq:Yukawa}
\end{align}
where $ X $ is the spurion field whose $F$-term breaks supersymmetry. The Yukawa is then:
\begin{align}
y_a= \frac{F}{M^2},
\label{eq:Yukawa1}
\end{align} 

This type of operator was  studied in \cite{Ibe:2010ig}, \cite{Dobrescu:2010mk}, and can provide the dominant contribution to down-type quarks masses. In the model of  \cite{Dobrescu:2010mk} for example, it is generated through loops of superpartners.  However, in the MMRSSM it is not generated through loops of particles present below the messenger scale, and 
In order to generate it, it is necessary to enlarge again the messenger sector. We can, for example introduce new link superfields  $ X_u, $ and $ X_d $ with the same gauge numbers of $H_u$ and $ R_d $ respectively, but  with different $ R$-charges:$ X_u$ has $R$-charge $2,$ while $X_d$ has R-charge $0$. They couple to visible sector and messenger fields through  superpotential couplings of the form:
\begin{align}
W_{y_a}= M_X X_u X_d+ y_1 X_d H_u X + y_2 X_d L_a e_a^c
\end{align}
When $X_d$ is integrated out at tree level, it yields the operator of equation \eqref{eq:Yukawa}. However, it also yields a tree level contribution to the Higgs soft mass squared. This last contribution can be made smaller than the gauge mediated Higgs soft mass  by choosing $y_1$ to be small. Then, to generated a large enough Yukawa coupling, the SUSY breaking scale must be rather low. For example, to generate the electron Yukawa ($a=e$), assuming $M_X$ to be of the same order as the messengers and setting the gaugino at $\sim 1$ TeV, the bound is given by:
\begin{equation}
\Lambda \lesssim 10^3 \text{TeV} \; \;.
\end{equation}
In this context, generating the $\tau$ Yukawa would require making $M_X$ smaller.  Another possibility is to generate the Yukawa at one loop by coupling $X_{u,d}$ to $X$ via the $S$ field of equation \eqref{eq:musuperpotential}:
\begin{align}
W_{y_a}= M_X X_u X_d+ y_1 X_d L_a l^c_a+ y_2 H_u X_d \bar S+y_3 X_u X_d S,
\label{eq:Yukawa3}
\end{align}
where $ S $  and $ \bar S $ are the link fields of eq.(\ref{eq:musuperpotential}), and  we assume $ M_X \sim M_S \sim \sqrt{F}.$
Then, the effective following operator  receives  contribution at one loop:
\begin{align}
c \int{\frac{d^4 \theta}{M_S^2 M_T^2} D^2 \left(X^{\dagger} X\right) H_u^{\dagger} L_a l^c_a},
\label{eq:Yukawa5}
\end{align}
with
\begin{align}
c \sim  \frac{ \lambda {\tilde \lambda}^2 y_1 y_2 y_3  }{16 \pi^2} \; \;.
\label{eq:Yukawa4}
\end{align}
but there is no contribution to the Higgs soft mass at the same order. This, taking the gaugino at $1$ TeV, will give a yukawa coupling of the order of:
\begin{equation}
y_a \sim 10^2 \left( \frac{1 \text{TeV}}{\Lambda} \right)^2.
\end{equation}
In this way, a Yukawa coupling for the $\tau$ can be accommodated, but requires a low SUSY breaking scale.
\section{\label{pheno}Phenomelogy}
\subsection{MMRSSM at the LHC}
The Dirac nature of  the gauginos
is one of the most distinctive  phenomenological aspects  of  models with a continuous $ R $-symmetry. It could provide a way to distinguish this type of models from  
the standard SUSY scenario where the gauginos are Majorana fermions. The phenomenology  of Dirac gauginos  versus Majorana gauginos has been examined in \cite{Freitas:2009dp}.
In addition,  the phenomenology of the MRSSM  Higgs sector has been recently discussed in \cite{Choi:2010an}, and it has been  noted that the inert doublet/doublets
\footnote{In our model and in the SOHDM there is just
one inert doublet$,H_d,$ while in the MRSSM a couple $R_u,$ and $ R_d.$}
with $ R $-charge $2$ can provide interesting 
signatures.

\par
 The  MMRSSM  has additional distinguishing features because of the identification of the $ U(1)_R  $ with a  lepton number. Because the model does not respect the standard R-parity, the lightest superpartner (LSP) is unstable, as in R-parity breaking models. Since most superpartners are charged under the lepton number $a$,  their decay chain will typically produce many leptons.  Moreover, in the MMRSSM the LSP is always the gravitino. As a result, to study the typical decay chain we should look at the  next lightest SUSY particle (NLSP).
In pure $D$-term SUSY breaking scenarios, the right handed sleptons are typically the lightest particles after the gravitino. Therefore the right handed stau $ \tilde \tau_ R$ is the NLSP. 
When  $a=e$ or $ a=\mu $, then there are two body decays for $ \tilde \tau_ R$:
 \begin{align}
    \tilde \tau^{\pm} &\rightarrow \nu_{\tau} a^{\pm}, \\
    \tilde \tau^{\pm} &\rightarrow \nu_{a} \tau^{\pm},
    \end{align}
 Typical decay chains will then contain jets, electron (or muon), plenty of tau's (up to 4), and missing energy from the neutrinos.  This kind of signature is also present in $R_P$ violating models with $\tilde \tau$ LSP (see  \cite{Desch:2010gi}).
\par
If instead the NLSP is the lightest gaugino $ \chi_1^0$, the situation is a little bit different.  
The possible $ \chi_1^0 $  decay modes are:
    \begin{align}
    \chi_1^0& \rightarrow Z^0 \nu_a, \\
    \chi_1^0& \rightarrow W^{\pm} l_a ^{\mp},
    \end{align} 
    which are driven by the mixing with the neutrino $a$.
Again, the same phenomenology  can be seen in the context of a $R_p$ parity violating models.
  \par
In summary, the MMRSSM phenomenology is similar to the phenomenology of  models with $ R_p $ violation. However, there are still important differences. First, we can exploit the Dirac nature of the gauginos by looking for example at same sign dileptons signatures.
Secondly, as it has been discussed in section 3, the MMRSSM can tolerate a larger level  $R_p$ parity violation than in the standard $R_p$ violating models due to the absence of constraints from neutrino physics.
Indeed, in the typical  $ R_P $ violating  scenario all decay chains end  in the LSP or in the NLSP,  whose decay modes are driven by the trilinear $ R_p$ breaking couplings.
Instead, in the MMRSSM the trilinear coupling  can be significantly  larger and this can lead to a distinctive phenomelogy.
 The most promising channels are  the decay of the right handed sbottom and left handed stop \footnote{ Right handed and Left handed sfermions don't mix in $ R$ symmetric models}, which are
the following:
\begin{align}
\tilde b_R \rightarrow b \nu_a \\
\tilde t_L \rightarrow  t l_a \
\end{align}
These decay modes can have significant branching ratios, and therefore can lead  to interesting signatures typical of  leptoquark phenomenology.
\par
Therefore, the MMRSSM possesses a quite distinctive phenomelogy at colliders that would be interesting to explore further in a future work.

\section{Conclusions and Outlook}
Supersymmetric models with Dirac gauginos are an interesting alternative to the more common MSSM scenario where gauginos are Majorana. They might help in making the gaugino naturally heavy, they can have a $U(1)_R$ symmetry which helps with flavour bounds, and present a different framework for SUSY breaking and mediation. In this work we have presented a supersymmetric model with a $U(1)_R$ symmetry in which the down-type quark and leptons get their mass from the vev of a sneutrino. The usual down-type Higgs is kept to cancel anomalies and give mass to gauginos, but is an inert doublet.  This allows for a  reduced particle content in the Higgs sector of such models which would otherwise require the addition of two new doublets. It is possible to realize this scenario because the $U(1)_R$ symmetry of the model can be identified with a lepton number. 

There are various constraints on such a setup and the main goal of this paper was to examine them and determine if the model is viable. The first constraint comes from electroweak precision measurements. There are two new sources of contributions to electroweak precision observables in this model. First, because one the lepton doublet mixes with the gauginos, its coupling to the $W$ and $Z$ are modified, putting an lower bound on $\tan \beta$. Secondly, the down-type Yukawa coupling can now give tree-level contributions to some electroweak observables, which give an upper bound on $\tan \beta$. We found that nevertheless, a large fraction of the possible parameter space is still viable. 

Another source of constraint on this model comes from the fact that the $R$-symmetry is not an exact symmetry and will be broken by the gravitino mass. Such a breaking will be communicated to the visible sector by anomaly mediation, if nothing else. This, in our scenario, will break lepton number and induce a mass for a neutrino. In order for this neutrino to be light enough, the anomaly mediation contribution to soft SUSY breaking terms must be very subdominant. This point towards a scenario of low scale susy breaking mediation for our model. This will have consequences, for example, for the resolution of the $\mu/B_{\mu}$ problem, which we also started exploring in this work. 

Finally, the gravitino in this model is unstable. It decays to a neutrino and a photon, and it's abundance is constrained by the observation of gamma-ray lines from the galactic center, and from diffuse photon background. We found that this constraint could be satisfied by invoking a very low reheating temperature, below the SUSY scale. 

There remain many issues to explore in this class of models. For example, it would be interesting to tie in our scenario with a concrete SUSY breaking mediation model. We could then explore issues such as fine-tuning, get a clearer picture of the expected phenomenology, and see how well the gauge couplings unify. It would also be interesting to see if the Higgs LEP bound could be ameliorated through an NMSSM-like model. 

The other aspect which we did not touch upon concern flavour. We singled out one flavour of lepton whose associated lepton number we identified with the $U(1)_R$. It is also important in our model that the charged lepton Yukawa matrix be very nearly diagonal, with neutrino mixing put in the Majorana neutrino mass terms. A very natural question to ask is then how flavour physics fits in, and how easy  it is to realize our requirements in various flavour models.

Finally, there remain unanswered question regarding the cosmological consequences of our model. In particular, our model does not contain a viable dark matter candidate as the gravitino abundance is constrained to be very low. Adding a dark matter sector and exploring possibilities for baryogenesis in the light of the low reheating temperature constraints will be required to make this framework more realistic from the cosmological point of view.
\section*{Acknowledgements}
CF would like to thank Enrico Bertuzzo for valuable discussions.
\par
 This work was supported  in part by the Natural Sciences and Engineering Research Council of Canada (NSERC).

\end{document}